\documentstyle[floats,aps,psfig,prb]{revtex}
\begin{document}
 
\flushbottom
 
\twocolumn[\hsize\textwidth\columnwidth\hsize\csname @twocolumnfalse\endcsname

\title{
Superconductor-insulator quantum phase transition in a single 
Josephson junction
}
\author{Carlos P. Herrero$^1$ and Andrei D. Zaikin$^{2,3}$}
\address{$^1$Instituto de Ciencia de Materiales,
         Consejo Superior de Investigaciones Cient\'{\i}ficas,
         Campus de Cantoblanco, 28049 Madrid, Spain \\
         $^2$Forschungszentrum Karlsruhe, 
         Institut f\"ur Nanotechnologie, 76021 Karlsruhe, Germany\\
         $^3$I.E. Tamm Department of Theoretical Physics, P.N. Lebedev
         Physics Institute, Leninski av. 53, 117924 Moscow, Russia }
\date{\today}
\maketitle

\begin{abstract}
The superconductor-to-insulator quantum phase transition in resistively shunted
Josephson junctions is investigated by means of path-integral Monte Carlo
simulations. Our numerical technique allows to directly access the (previously
unexplored) regime of the Josephson-to-charging energy ratios $E_J /
E_C$ of order one. Our results unambiguously support an earlier
theoretical conjecture, based on 
renormalization-group calculations, that at $T \to 0$ the dissipative phase 
transition occurs at a universal value of the shunt resistance $R_S =
h/4e^2$ for all values  $E_J / E_C$. On the other hand,
finite-temperature effects are shown to turn
this phase transition into a crossover, which position depends 
significantly on $E_J / E_C$, as well as on the dissipation strength and on 
temperature. The latter effect needs to be taken into account in order
to reconcile earlier theoretical predictions with recent experimental results.
\end{abstract}


\vskip2pc]
 
\narrowtext 

\section{Introduction}
Mesoscopic Josephson junctions are well known to exhibit a variety
of intriguing phenomena which are of primary importance both from a fundamental
point of view, as well as for various applications including quantum-state
engineering with electronic devices. Among these are macroscopic quantum tunneling with
dissipation,\cite{CL} Coulomb blockade,\cite{AL,SZ,gr92} macroscopic quantum
coherence, and dissipative quantum phase transitions.\cite{SZ} Recent progress
in nanolithographic techniques allows one to routinely fabricate ultrasmall tunnel 
junctions with capacitances $C$ as low as $10^{-15} - 10^{-16}$ F, and
to perform detailed experimental studies of various features related
to the above phenomena. 

In the course of these studies it was realized -- both theoretically
and experimentally -- that the observed properties of the system 
may crucially depend on the nature of the effective electromagnetic environment 
coupled to a mesoscopic junction. One of the most remarkable consequences of this
dependence for Josephson junctions (JJ) is the quantum ($T=0$) superconductor-to-insulator 
phase transition (SIT) driven by dissipation. The latter is controlled, 
e.g., by the magnitude of the ohmic shunt resistance $R_S$ of the 
external leads. This 
quantum phase transition was predicted by Schmid \cite{sc83} and subsequently
studied in Refs. \onlinecite{bu84,gu85b,FZ} (see also Refs. \onlinecite{SZ,Weiss} for
an extensive review of this and later theoretical activity). Thus, at low 
temperatures the supercurrent in mesoscopic JJs can be maintained only provided
that quantum fluctuations of the Josephson phase are suppressed by dissipation. 
If dissipation is not strong enough, quantum fluctuations wash out the 
Josephson effect and no supercurrent
can flow across the system. Alternatively, one can interpret this SIT as a destruction
of Coulomb blockade for Cooper pairs by quantum fluctuations of the charge in an 
external resistor $R_S$. Experimentally this quantum dissipative phase
transition for single resistively shunted JJs was studied in 
Refs. \onlinecite{ya97,sh96,pe99,pe01}.
The results of these experiments are qualitatively consistent with the above 
physical picture.

Similar experimental studies have been also performed for JJ arrays and 
chains.\cite{Clarke,ta00,ha00}
In that case an interplay between short and long-range quantum fluctuations 
of the superconducting phase in the presence of dissipation yields a nontrivial
 low-temperature phase diagram.\cite{PZ,Bobbert,Saro} 
A quantum dissipative phase transition was also discussed in the case
of ultra-thin homogeneous superconducting wires.\cite{ZGOZ,Tinkham} 

It is worth to point out
that the above physical picture is not restricted to superconducting 
systems only. For instance,
it is well known that the problem of a quantum resistively shunted JJ is equivalent to 
that of a quantum particle diffusing in a periodic potential
coupled to a dissipative environment. In this case, the phase transition from diffusion
to localization occurs upon increasing the coupling strength to a dissipative 
Caldeira-Leggett bath.\cite{sc83,bu84,gu85b,FZ,Weiss}  A formally
identical Lagrangian also describes tunneling of electrons in a Luttinger liquid; 
see, e.g., Ref. \onlinecite{Weiss}. Similar physics was discussed for normal 
metallic conductors.\cite{PZ91} 
Thus, even though below we will specifically address the case of a resistively 
shunted JJ, our results can also be applied in other physical situations.   

According to the existing theoretical picture,\cite{SZ,Weiss} at $T=0$ quantum 
localization of the Josephson phase should occur as the shunt resistance 
becomes equal to the quantum resistance 
unit, $R_S=R_q=\pi \hbar /2e^2 \simeq 6.5$ k$\Omega$, {\it independently  }
 of the strength of the
Josephson coupling $E_J$. In the limit of small Josephson energies (as compared
to the effective charging energy of the junction, $E_C=e^2/2C$) 
this conclusion can be justified  within a perturbative renormalization
group analysis. Such an analysis can then be extended
to the limit of large $E_J\gg E_C $ by means of a duality transformation between the
phase and the charge.\cite{SZ,Weiss} Since in both limits one obtains an 
$E_J$-independent phase boundary at $R_S=R_q$, it is reasonable to 
{\it conjecture} that its position remains unchanged
for all values $E_J/E_C$ including the regime of experimental interest $E_J \sim E_C$. 

Although this conjecture can further be supported by a number of qualitative 
arguments, as well as by the existence 
of a self-duality point,\cite{sc83,Weiss} the structure of the phase diagram 
for moderate values $E_J/E_C$ still needs to be rigorously verified. 
Moreover, the results of recent experiments
\cite{pe99,pe01} could be interpreted as {\it contradicting} the above conjecture.
 In these works, the samples
with sufficiently large $E_J/E_C \gtrsim 7 \div 8$ were found to be superconducting 
even for 
shunt resistances $R_S$ substantially higher than $R_q$. This observation could 
suggest that the true phase boundary should depend not only on the amount of 
dissipation in the system, but also on the
ratio $E_J/E_C$. A similar conclusion could be reached from the results reported 
in Refs. \onlinecite{ya97,ta00}.

All these developments motivated us to perform an additional theoretical investigation of the
dissipative phase transition in a single resistively-shunted superconducting junction, 
at moderate values of the Josephson coupling energy  $E_J \sim E_C$. Since in this 
range there exists no small parameter in the problem, it can hardly be rigorously 
investigated by analytical methods. 
Therefore, in this paper we analyze the problem numerically by means of
path-integral (PI) Monte Carlo (MC) simulations. This method allows us to 
quantitatively study the effect of quantum fluctuations of the 
Josephson phase, depending on the dissipation parameter  $\alpha=R_q/R_S$ and 
the ratio $E_J/E_C$ in the interesting parameter range. 

Our main conclusions can be summarized as follows: (i) Our detailed MC analysis 
unambiguously supports an earlier conjecture that in
the zero temperature limit the superconductor-to-insulator phase transition always 
occurs at the dimensionless dissipation strength $\alpha =1$, independently of the ratio  
$E_J / E_C$, (ii) at nonzero
temperatures this phase transition is substituted by a {\it crossover}, 
which position depends on the ratio $E_J / E_C$, as well as on temperature $T$ 
and dissipation strength $\alpha$, and (iii) at
$T \to 0$ this crossover line approaches the phase transition line  $\alpha =1$.
These observations allow to fully reconcile the existing theoretical picture of the
dissipative phase transition in
a single resistively shunted JJ with the experimental results.\cite{pe99,pe01}

\section{Quantum dissipative phase transition}

We proceed within the standard path integral formulation of the problem outlined 
in Ref. \onlinecite{SZ}. The grand partition function of the system ``JJ+shunt''
can be expressed as a path integral over the Josephson phase $\phi$
\begin{equation}
Z \sim \int {\cal D} \phi \exp (-S[\phi]/\hbar )  \, ,
\label{Z}
\end{equation}
where $S$ is the effective action 
\begin{eqnarray} 
 S[\phi] = \int_0^{\beta}\left[\frac{\hbar^2}{16 E_C} \left( \frac{d \phi}{d\tau} \right)^2 
 -  E_J \cos \phi(\tau)\right] d\tau  +  \nonumber  \\ 
  +  \frac{\alpha \hbar}{8\beta^2} \int_0^{\beta} d\tau \int_0^{\beta} d\tau'
      \frac{ \left[\phi(\tau) - \phi(\tau')\right]^2  } 
          {\sin^2 [ \pi (\tau -\tau' )/\beta ]}  
   \label{action}
\end{eqnarray} 
and $\beta \equiv \hbar /k_BT$. The first and second terms in $S[\phi]$ account
 respectively for the charging
and Josephson contributions. The third -- nonlocal in time -- term
describes dissipation produced by an ohmic resistor.
An additional dissipative contribution to the action, due to tunneling quasiparticles, is 
usually negligibly small in the interesting limit of energies/temperatures well below
the superconducting gap. Hence, in what follows this contribution will be 
ignored for the sake of simplicity.

MC simulations with this effective action have been carried out
 by the standard discretization of the quantum
paths into $N$ imaginary-time slices.\cite{su93}
In order to maintain the accuracy of the calculation, as the
 temperature is lowered, the Trotter number $N$ has been
 increased proportionally to the imaginary time $\beta $.
In our simulations we have taken $N = 4 \beta E_C / \hbar$. 
This choice was proven to be sufficient 
in order to reach the necessary convergence of the calculated quantities:
 Repeating the calculations for some data
points with  larger $N$ did not change our results. 
We have employed the Metropolis sampling to 
carry out PI MC simulations at temperatures down to
$k_B T = E_C / 125$. 
A simulation run consists of successive MC steps, being updated
all path-coordinates in each step.
 For each set of parameters ($\alpha$, $E_J/E_C$, $T$),
we generated $5\times10^5$ quantum paths for the calculation of
ensemble-averaged values. 
More details on this kind of MC simulations can be
found elsewhere.\cite{he99}

In order to quantify the quantum delocalization of the phase we consider the 
mean-square fluctuations of the quantum paths
\begin{equation}
\langle (\delta \phi)^2\rangle = \langle (\phi(\tau) - \bar{\phi})^2 \rangle,
\label{dphi}
\end{equation}
where $\bar{\phi} = \beta^{-1} \int_0^{\beta} \phi(\tau) d\tau$
is the mean value of the phase for a given path. The quantity (\ref{dphi})
was calculated by means of PI MC simulations with the aid of Eqs. (\ref{Z}) 
and (\ref{action}).

In Fig. 1(a) we present $\langle (\delta \phi)^2\rangle$ as a function of $\alpha$
for $E_J/E_C = 2$. Different symbols correspond to different
temperatures; from top to bottom: $k_B T / E_C$ = 0.02, 0.05,
and 0.11. One observes that the phase delocalization increases
 as the temperature is lowered.  At $\alpha = 1$ 
one also observes a change of the slope in the data
points. This change becomes more and more pronounced as the temperature is lowered,
indicating the expected tendency to a vertical line at $\alpha < 1$ and $T \to 0$.
Unfortunately the noise in the calculated values
increases as $T$ is reduced, especially for $\alpha < 1$,
where one expects an insulating regime ($\phi$ delocalized).
For $\alpha > 1$, however, the quantum paths obtained in the
MC simulations are merely confined to a single potential well, and
the phase $\phi$ is localized (superconducting regime).
A similar picture is found for all other values of $E_J/E_C$ at which our simulations
have been performed. The results obtained for $E_J/E_C=0.75$ are presented
in Fig. 1(b) for comparison. In agreement with intuitive expectations, one finds that
the phase fluctuations increase with decreasing ratio $E_J/E_C$. However, 
the phase transition point $\alpha = 1$ remains insensitive to this ratio in
all cases.
 
\begin{figure}
\vspace*{-1.5cm}
\centerline{\psfig{figure=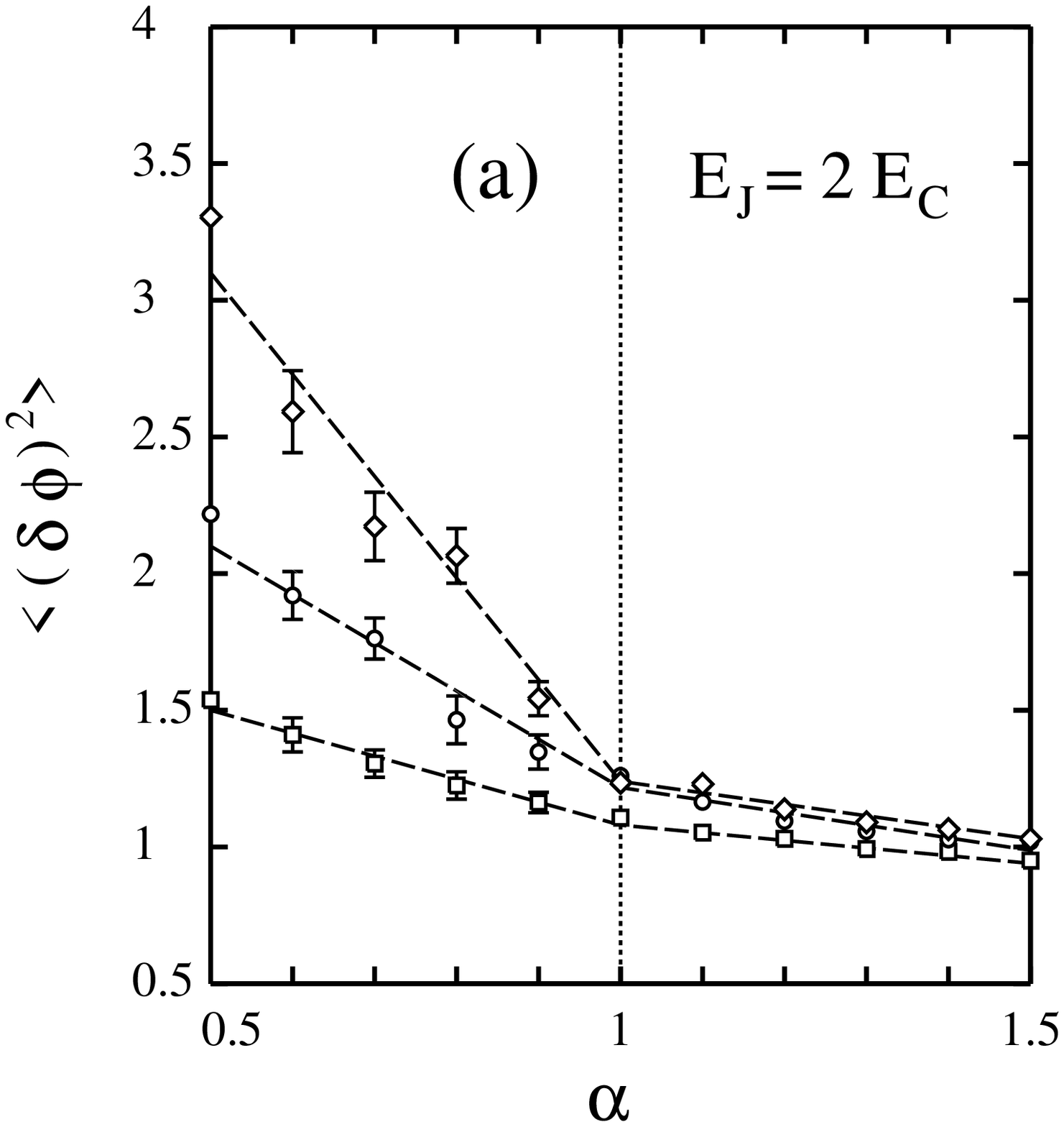,height=9.0cm}}
\vspace*{-3.5cm}
\end{figure}
 
\begin{figure}
\centerline{\psfig{figure=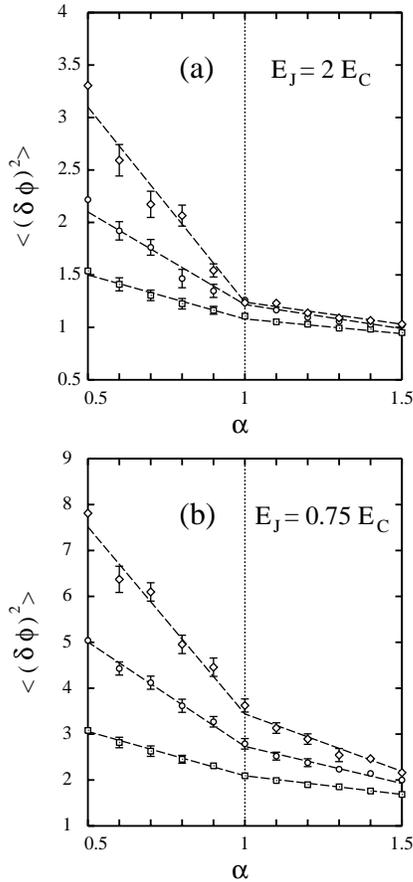,height=9.0cm}}
\vspace*{-1.5cm}
\caption{
Mean-square displacement $\langle (\delta \phi)^2\rangle$
as a function of the dissipation strength
$\alpha$ for (a) $E_J = 2 E_C$ and (b) $E_J = 0.75 E_C$ at three different
temperatures: Squares, $k_B T = 0.11 E_C$;
circles, $k_B T = 0.05 E_C$; diamonds, $k_B T = 0.02 E_C$.
} \label{f1} \end{figure}

In Fig. 2 we show the average mean-square displacement
$\langle (\delta \phi)^2\rangle$ versus $\alpha$ for several values of the 
ratio $E_J/E_C$ at a temperature low as compared to the charging
energy: $k_B T = 0.02 E_C$.
Values of $E_J/E_C$ increase from top to bottom: 0.2, 0.5, 1, and 2.
As expected, for a given dissipation strength $\alpha$, the phase becomes 
``more localized'' as the ratio $E_J/E_C$ increases, since the effective potential
 barrier for the phase increases as well.  
Fig. 2 also demonstrates that stronger suppression of the phase
fluctuations $\langle (\delta \phi)^2\rangle$ is always obtained when the 
dissipation strength $\alpha$ increases for a given value of $E_J/E_C$. 
Again in all cases we observe that the change of the slope in the
data points for $\langle (\delta \phi)^2\rangle$ occurs exactly at $\alpha = 1$, 
indicating the existence of the phase transition at this point.

\begin{figure}
\vspace*{-1.5cm}
\centerline{\psfig{figure=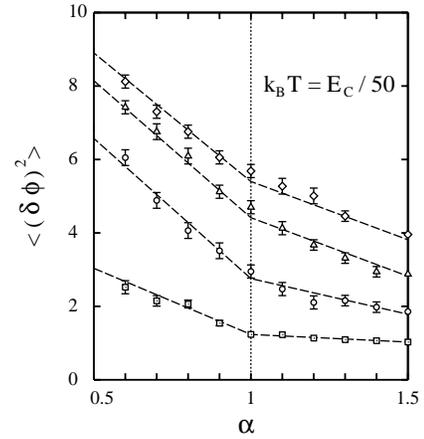,height=9.0cm}}
\vspace*{-1.5cm}
\caption{
Mean-square displacement $\langle (\delta \phi)^2\rangle$
as a function of the dissipation strength
$\alpha$ for $k_B T = 0.02 E_C$.
Results are given for several values of the
ratio $E_J / E_C$. From top to bottom:
$E_J / E_C$ = 0.2, 0.5, 1, and 2.
Error bars, if not shown, are of the order of
the symbol size.
} \label{f2} \end{figure}

Further information can be obtained by investigating the dependence
of $\langle (\delta \phi)^2\rangle$ on temperature.
As expected, in our simulations we observe that for all  values of
$E_J/E_C$ and $\alpha$, the quantum delocalization of $\phi$ increases as the
temperature is lowered. In Fig. 3(a) we have plotted the temperature
dependence of $\langle(\delta \phi)^2\rangle$ for $E_J/E_C = 2$.
Different symbols represent several $\alpha$ values,
which increase from top ($\alpha$ = 0) to bottom ($\alpha$ = 1.2).
At low $T$ ($E_C \gtrsim 10 k_B T$), for all values of
$\alpha$ one finds that $\langle (\delta \phi)^2\rangle$ follows a power-law 
dependence on temperature, as shown by the dashed lines in Fig. 3(a).
In particular, in the dissipationless limit $\alpha$ = 0, it is reasonable
to expect the quantum paths to be in a diffusive regime, so that
$\langle(\delta \phi)^2\rangle \propto \beta$. This is indeed confirmed by 
our calculations.
In the presence of dissipation the phase diffusion slows down, as indicated by
a decrease in the slope of the dashed lines in Fig. 3(a). From this plot we 
find $\langle(\delta \phi)^2\rangle \propto \beta^\gamma$, with an exponent 
$\gamma $ smaller than unity for $\alpha >0$.
Again, the localization phase transition at $\alpha =1$ is clearly observable 
in Fig. 3(a), since essentially no diffusion of the phase takes place at 
$\alpha \gtrsim 1$ and sufficiently large $\beta$.
Someting similar occurs for other ratios $E_J / E_C$, as shown in Fig. 3(b)
for $E_J / E_C = 0.75$. The main difference with the previous case is that 
the low-temperature regime $\langle(\delta \phi)^2\rangle \propto \beta^\gamma$
is reached at lower $T$ ($E_C \gtrsim 30 k_B T$).
 
In order to find the parameter $\gamma$, for each $\alpha$ we numerically evaluated
the logarithmic derivative 
 $d \ln \langle (\delta \phi)^2\rangle /d\ln \beta  \equiv \gamma$.
The values $\gamma$ obtained in this way are shown in Fig. 4 as a function
of $\alpha$ for $E_J/E_C$ = 2.  These results clearly indicate a linear dependence
of $\gamma$ on $\alpha$ of the form $\gamma = 1 - \alpha$. 
Thus, from our numerical analysis we can conclude that at sufficiently
low temperatures and $\alpha < 1$ the phase diffusion is described by the formula
\begin{equation}
\langle(\delta \phi)^2\rangle \propto \beta^{1-\alpha}.
\label{dif}
\end{equation}
This dependence turns out to apply for all values $E_J/E_C$ used in our simulations.
From a linear fit to our data points in Fig. 4
we find the transition point at $\alpha = 1.02 \pm 0.04$.
 
\begin{figure}
\vspace*{-2.0cm}
\centerline{\psfig{figure=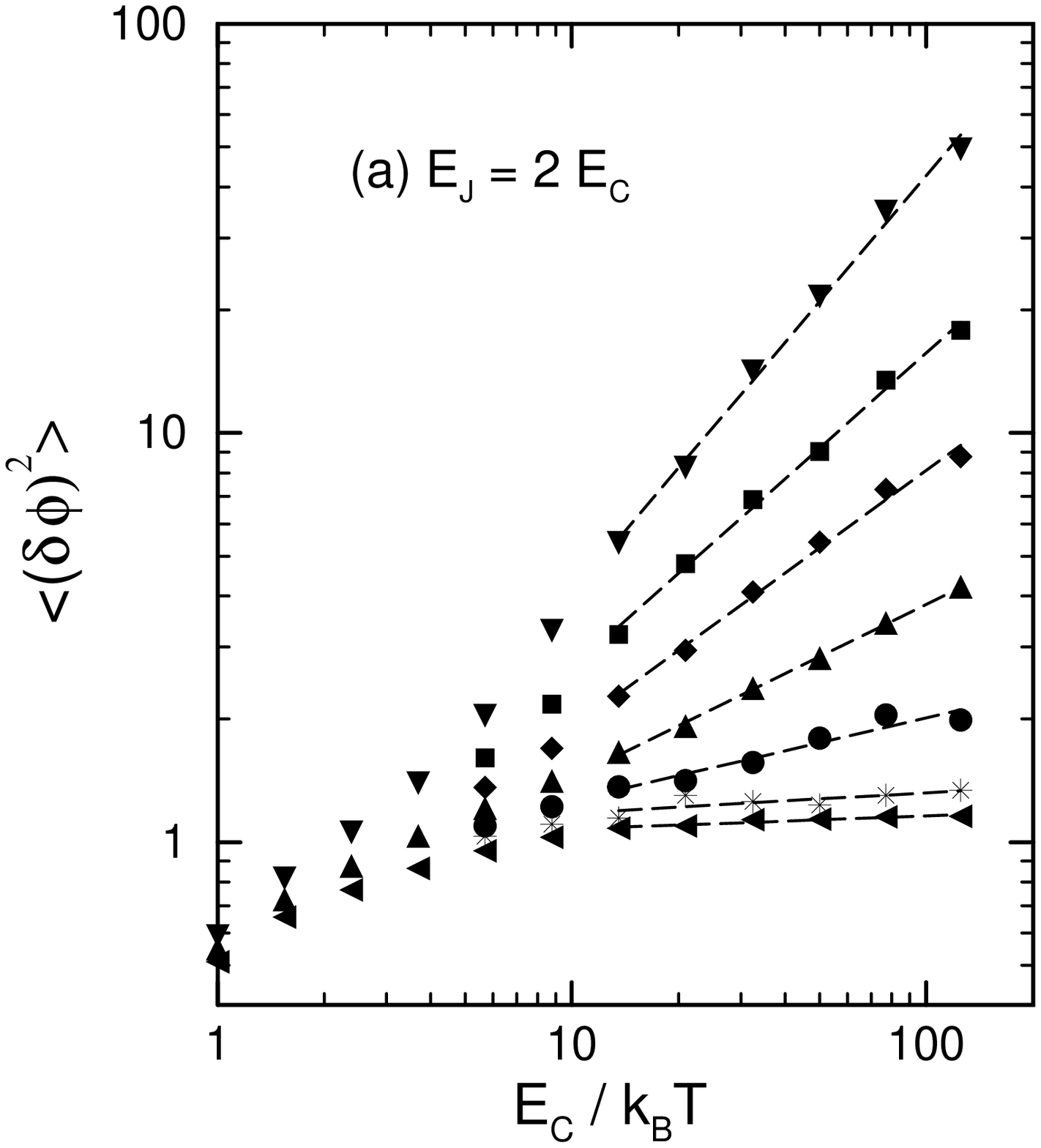,height=9.0cm}}
\vspace*{-3.1cm}
\end{figure}
 
\begin{figure}
\centerline{\psfig{figure=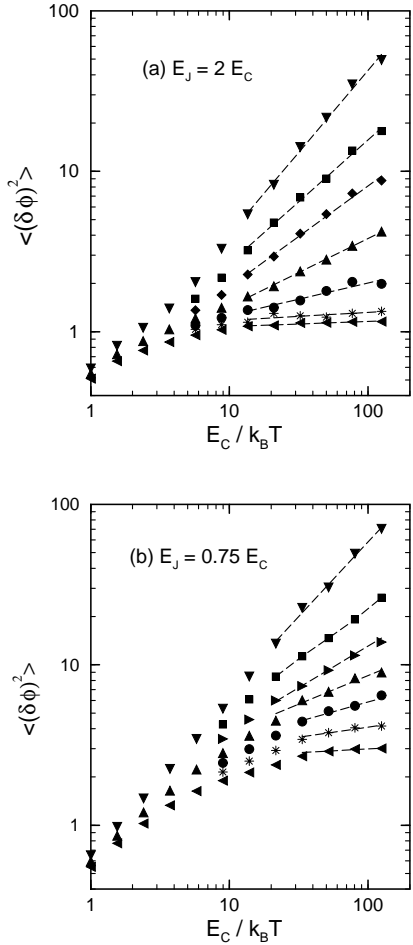,height=9.0cm}}
\vspace*{-0.5cm}
\caption{
Temperature dependence of
$\langle (\delta \phi)^2\rangle$ for
(a) $E_J = 2 E_C$ and (b) $E_J = 0.75 E_C$ at different values of
$\alpha$.   From top to bottom:
$\alpha$ = 0, 0.2, 0.4, 0.6, 0.8, 1, and 1.2.
Dashed lines indicate the low-temperature trend of the data points.
} \label{f3} \end{figure}

The prefactor in Eq. (\ref{dif}) depends on both $E_J$ and  $E_C$.
In the limit $E_J\gg E_C$, one can demonstrate  with the aid of the instanton 
technique \cite{SZ} that this prefactor is proportional to $\Delta^{1-\alpha}$,
where
\begin{equation}
\Delta =16 \left(\frac{E_JE_C}{\pi}\right)^{1/2} \left(\frac{E_J}{2E_C}\right)^{1/4}
\exp \left( -\sqrt{8E_J/E_C} \right)
\label{bw}
\end{equation}
is the bandwidth without dissipation.  For $E_J \sim E_C$ this
equation does
not apply anymore, but the qualitative trend remains the same: For a
given value of $E_C$ the prefactor increases monotonously with decreasing  $E_J$;
cf. Figs. 3(a) and 3(b).

To conclude this part of our analysis we emphasize again that -- even though
we present here our numerical results for the energy ratios $E_J / E_C$ = 2 
and 0.75 -- 
similar behavior is observed for other values of the Josephson coupling energy. 
In particular, we have performed detailed
MC simulations also for $E_J / E_C$ = 0.25 and 3, and found essentially the same
behavior as the one discussed above. In all cases, from our numerical data we obtained 
unambiguous indications of the quantum localization phase transition at $\alpha = 1$. 
 
\begin{figure}
\vspace*{-1.5cm}
\centerline{\psfig{figure=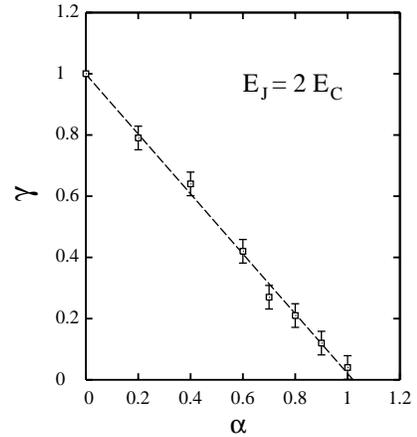,height=9.0cm}}
\vspace*{-1.5cm}
\caption{
Low-temperature exponent $\gamma$ as a function of $\alpha$
for $E_J = 2 E_C$. The dashed line is a linear fit to the data
points.
} \label{f4}  \end{figure}

\section{Effect of temperature}

Now let us see how the above physical picture is modified at nonzero $T$. Since
any experiment is performed at a finite temperature, it is important to find out
if the SIT can be observed under such conditions. 

To begin with, we recall the 
argument \cite{SZ} according to which superconductivity in Josephson junctions
can be observed even for  $\alpha < 1$, provided that the phase diffusion is slow enough,
in the sense that the characteristic delocalization time $\tau_{\rm del}$ for the phase $\phi$ 
exceeds the time of experiment. This situation can easily be achieved in the
limit  of very large $E_J \gg E_C$. In that case the time $\tau_{\rm del}$ is 
exponentially large, and no delocalization effects can be detected in a real
experiment. This argument imposes serious limitations on the observation
of a SIT for sufficiently large values of  $E_J$ even at $T \to 0$. 
In Ref. \onlinecite{pe99}
this argument was extended taking into account the accuracy of 
the voltage measurements. The authors \cite{pe99} argued that for
their experiment  $1/\tau_{\rm del}$ should be compared with the
quantity 
$eV_{\rm min}$
rather than with the typical experimental time, where $V_{\rm min}$ is the minimum 
voltage detectable in the experiment. They also noticed that further limitations
can occur due to temperature effects.

To explore the latter possibility we first notice that according to our result
(\ref{dif}), at any nonzero temperature quantum fluctuations do not spread the phase
$\phi$ to infinity even for $\alpha < 1$. The phase does not simply have ``enough
time'' to diffuse, and  $\langle (\delta \phi)^2\rangle$ remains finite though increasing with 
the inverse temperature $\beta$. Thus, at nonzero $T$ and not very small $E_J$ one might 
expect to observe a nonvanishing supercurrent even at small dissipation. This
conclusion might appear paradoxical. One can argue that, if quantum fluctuations
of the phase yield suppression of superconductivity already at $T=0$ 
(and $\alpha < 1$), at nonzero temperatures this suppression can only increase further,
because of an additional effect of thermal fluctuations. 

In order to understand why this conclusion might not be quite correct, it is instructive to
analyze the behavior of the (quasi-)charge variable\cite{SZ} canonically conjugate
to the Josephson phase $\phi$. As discussed above, at $T=0$ and  $\alpha < 1$
the quasicharge is localized, i.e. Cooper pairs cannot tunnel across the junction due
to Coulomb blockade and, hence, the junction behaves as an insulator. At nonzero $T$
this behavior persists as long as the temperature remains much smaller than the effective
Coulomb gap for Cooper pairs. At  $E_J \ll E_C$ this Coulomb gap is large ($\approx E_C$), 
while in the opposite limit $E_J \gg E_C$ it turns out to be exponentially small:\cite{SZ}
\begin{equation}
\Delta_r = \Delta \left(\Delta \over\hbar \omega_p 
       \right)^{\alpha / (1- \alpha)} \, .
\label{tdelta}
\end{equation}  
Here $\omega_p=\sqrt{8E_J E_C}/\hbar$ is the plasma frequency and  
$\Delta$ was defined in Eq. (\ref{bw}). If the temperature becomes higher than the 
gap (\ref{tdelta}), $k_B T \gtrsim \Delta_r$,
Coulomb blockade for Cooper pairs (and, hence, the insulating behavior) is destroyed. This
implies that the quasicharge $Q$ gets
strongly delocalized due to thermal effects. Because of the uncertainty
relation 
\begin{equation}
\delta Q\delta \phi \gtrsim e    \; ,
\label{unc}
\end{equation} 
delocalization
of $Q$ in turn restricts fluctuations of the canonically conjugate variable -- 
the Josephson phase $\phi$ -- and as a result of that the superconducting behavior
can be partially restored.
The same scenario can be reformulated in the phase space. One just needs to compare the typical
inverse time during which the phase diffuses at a distance $\sim 2\pi$ with temperature. 
In the limit  $E_J \gg E_C$ this inverse time  $\hbar /\tau_{\rm del} \sim \Delta_r$,
and we arrive exactly
at the same condition $k_BT\sim \Delta_r$ for the crossover line between ``insulating''
and ``superconducting'' phases. Within logarithmic accuracy this crossover line agrees 
with the one found in Ref. \onlinecite{pe99}.

Although the above consideration appears to be sufficient at high Josephson energies 
($E_J \gg E_C$),
no quantitative conclusion can yet be drawn for moderate couplings $E_J \sim E_C$.
 In order to study this parameter range one can apply a simple variational 
ansatz,\cite{FZ} which leads to the self-consistency equations
\begin{eqnarray}
D & = & E_J\exp (-\langle \phi^2\rangle_{\rm tr}/2)     \nonumber  \\
\langle \phi^2\rangle_{\rm tr}  & =  & k_B T \sum_n[C(\omega_n/2e)^2+
\alpha |\omega_n | /2\pi +D]^{-1}   \; ,
\label{var}
\end{eqnarray}
where $\omega_n=2\pi n / \beta$ are the Matsubara frequencies. At $T=0$ these equations
have a nonzero solution for $D$ (which corresponds to superconductivity) only at $\alpha > 1$, 
whereas at nonzero temperature one can get a positive solution $D>0$ also for $\alpha < 1$.
By resolving these self-consistency equations at different $T$, one can qualitatively describe 
the above crossover for moderate values of  $E_J$.

This crossover can be studied quantitatively by PI MC simulations, using the
effective action given above in Eq. (\ref{action}).
With this purpose we have evaluated numerically the correlation function 
$\langle\phi (\tau )\phi (0)\rangle$ for different values of
$E_J / E_C$,  dissipation strength $\alpha$, and temperature $T$. After
Fourier transformation, this correlation function is directly related to the 
``Matsubara resistance'' 
$\tilde R(\omega_n) = |\omega_n| \langle\phi\phi\rangle_{\omega_n}/4e^2$,
 which yields the system resistance
after analytic continuation to real frequencies. 
This numerical analytic continuation is
a separate complicated problem, which will not be discussed here. Fortunately,
however, this procedure is not needed in order to establish the position
of the crossover line. 
 
\begin{figure}
\vspace*{-1.5cm}
\centerline{\psfig{figure=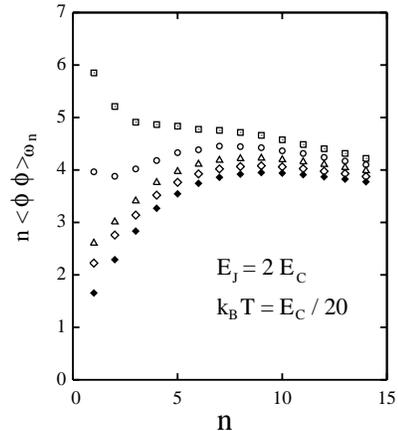,height=9.0cm}}
\vspace*{-1.5cm}
\caption{
Dependence of the ``Matsubara resistance''
$\tilde R(\omega_n)\propto |\omega_n|
\langle\phi\phi\rangle_{\omega_n}$
on $n=\omega_n \beta/2\pi$,
for $E_J / E_C = 2$ and $k_B T = E_C / 20$.
 Different symbols correspond to different
values of the dissipation strength. From top to bottom: $\alpha$ =
0.6, 0.7, 0.8, 0.9, and 1.
} \label{f5}  \end{figure}

Let us express the above correlation function in the form
\begin{equation}
|\omega_n|\langle\phi\phi\rangle_{\omega_n}=
\frac{|\omega_n|}{C(\omega_n/2e)^2+
\alpha |\omega_n | /2\pi +\tilde E_J}   \, ,
\label{cf}
\end{equation}
where $\tilde E_J$ is the effective (renormalized) Josephson coupling
energy. It is easy to see that in the low-frequency limit 
the behavior of the correlation function (\ref{cf}) is totally
different, depending on whether $ \tilde E_J$ remains nonzero 
(superconductivity) or is fully suppressed by quantum fluctuations
(insulating regime). In the first case, for sufficiently low frequencies
the function (\ref{cf}) should inevitably decrease with  $\omega_n$,
while in the second case this function should increase and saturate
at a finite value $ 2\pi /\alpha$ in the limit  $\omega_n \to 0$.
Thus, by studying the behavior of the function (\ref{cf}) it is 
possible to determine numerically from the PI MC simulations the position
of the crossover line at different temperatures.

In Fig. 5 we present our MC results for the function  
$|n|\langle\phi\phi\rangle_{\omega_n}$ at $E_J =2E_C$, $k_BT=E_C/20$,
and different values of $\alpha$. One observes that for sufficiently
small $n$ this function increases with decreasing  $n$ for $\alpha =0.6$
(upper curve), saturates for  $\alpha =0.7$, and decreases monotonously
for  $\alpha \geq 0.8$ (three lower curves). 
Thus, by studying the small-$n$ behavior of this function for different
$\alpha$ values and $E_J/E_C$ ratios, we arrive at the crossover
line for the temperature $k_BT=E_C/20$. Other temperatures
are treated analogously.

\begin{figure}
\vspace*{-1.2cm}
\centerline{\psfig{figure=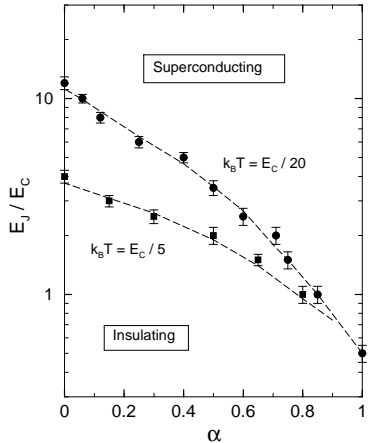,height=9.0cm}}
\vspace*{-1.5cm}
\caption{
Crossover lines between the insulating and superconducting regimes
at two different temperatures $k_B T = E_C / 5$ (squares)
and $k_B T=E_C / 20$ (circles).
Symbols are data points obtained from PI MC simulations.
Dashed lines are guides to the eye.
} \label{f6}  \end{figure}

 The resulting crossover lines are presented in Fig. 6 for two
temperatures: $k_BT=E_C/5$ and $E_C/20$. In full agreement with
the above qualitative considerations, one observes
that the position of the superconductor-insulator crossover
line is shifted towards larger values $E_J/E_C$ as the temperature
is lowered. The same trend is observed for all other temperatures used
in our simulations. Combining these results with those discussed
in the previous section, one arrives at the conclusion that 
the crossover line should approach the phase transition line $\alpha =1$
in the limit  $T\to 0$. We would also like to point out that the
position of the crossover line obtained within our MC analysis is
fully consistent with that found experimentally in Refs. 
\onlinecite{pe99,pe01}. It appears, therefore, that deviations
from the theoretical prediction for the phase boundary $\alpha =1$ 
observed in these experiments can be attributed to finite-temperature 
effects.\cite{FN}

In summary, the results of our MC simulations unambiguously demonstrate
that the quantum ($T = 0$) superconductor-insulator phase transition
in resistively shunted Josephson junctions
occurs at the value of the shunt  resistance $R_S = R_q$, irrespective of the
ratio $E_J / E_C$. For $\alpha < 1$, quantum diffusion 
of the Josephson phase $\phi$ yields a simple scaling dependence
$\langle (\delta \phi)^2\rangle \propto T^{\alpha -1}$.
Finite-temperature effects turn
the phase transition into a {\it crossover}, which position depends on the
ratio $E_J / E_C$, as well as on the dissipation strength $\alpha$, and on 
temperature. Our results are fully consistent with recent
experimental findings.\cite{pe99,pe01}

\acknowledgements

It is a pleasure to thank P.J. Hakonen and M.A. Paalanen for providing us with
the results of their work \cite{pe01} prior to its publication and for stimulating
discussions. This work is part of the CFN (Center for Functional 
Nanostructures) supported by DFG (German Science Foundation). 
C.P.H. acknowledges partial support by CICYT (Spain) under Contract
 No. BFM2000-1318. A.D.Z. acknowledges partial support by ULTI-III of the EU 
(HPRI-1999-CT-00050) and the hospitality
of the Low Temperature Laboratory of the Helsinki University of Technology,
where a part of this work has been performed.




\begin{references}
\bibitem{FN} The lowest temperature ($T \sim 80$ mK) achieved in the
experiments\cite{pe99} was considerably higher than $eV_{\rm min}/k_B$. Hence,
it appears that in these experiments temperature effects impose
more serious limitations as compared to those related to the
accuracy of the voltage measurements.
\end{references}
\end{document}